# Ultrafast reversal of the ferroelectric polarization

R. Mankowsky[1], A. von Hoegen[1], M. Först[1], A. Cavalleri[1,2]

[1]*Max Planck Institute for the Structure and Dynamics of Matter, 22761 Hamburg, Germany.*
[2]*Department of Physics, University of Oxford, Clarendon Laboratory, Oxford OX1 3PU, UK.*

The ability to manipulate ferroelectrics at ultrafast speeds has long been an elusive target for materials research. Coherently exciting the ferroelectric mode with ultrashort optical pulses holds the promise to switch the ferroelectric polarization on femtosecond timescale, two orders of magnitude faster compared to what is possible today with pulsed electric fields. Here, we report on the demonstration of ultrafast optical reversal of the ferroelectric polarization in $LiNbO_3$. Rather than driving the ferroelectric mode directly, we couple to it indirectly by resonant excitation of an auxiliary high-frequency phonon mode with femtosecond mid-infrared pulses. Due to strong anharmonic coupling between these modes, the atoms are directionally displaced along the ferroelectric mode and the polarization is transiently reversed, as revealed by time-resolved, phase-sensitive second-harmonic generation. This reversal can be induced in both directions, a key pre-requisite for practical applications.

The ferroelectric polarization is typically controlled with static or pulsed electric fields. Switching is in this case an incoherent process, with speed limited to hundreds of picoseconds by the nucleation and growth of oppositely polarized domains [1,2,3]. To overcome these limitations, attempts have been made to drive the ferroelectric mode coherently with light pulses. These strategies, which have been based either on impulsive Raman scattering [4,5,6,7] or direct excitation of the ferroelectric mode [8,9,10] using THz radiation, have not yet been completely successful.

Recent theoretical work [11] has analyzed an alternative route to manipulate the ferroelectric polarization on ultrafast timescales. It has been proposed that a coherent displacement of the ferroelectric mode could be achieved indirectly, by exciting a second, anharmonically-coupled vibrational mode at higher frequency. The underlying mechanism is captured by the following minimal model, which is illustrated in Fig. 1. Consider a double well energy potential along the ferroelectric mode coordinate $Q_P$ as shown in Fig. 1(b) (dashed black line)

$$V_P(Q_P) = -\frac{1}{4}\omega_P^2 Q_P^2 + \frac{1}{4}c_p Q_P^4.$$

Here, $\omega_p$ is the frequency of the ferroelectric mode for a fixed temperature $T$ below the Curie temperature $T_C$. Take then a second mode $Q_{IR}$ with frequency $\omega_{IR}$, described by the harmonic potential $V_{IR}(Q_{IR}) = \frac{1}{2}\omega_{IR}^2 Q_{IR}^2$ shown in Fig. 1(a), which is anharmonically coupled to the ferroelectric mode with a quadratic-linear dependence of the interaction energy $aQ_{IR}^2 Q_P$. The total lattice potential can then be written as

$$V(Q_P, Q_{IR}) = V_P(Q_P) + V_{IR}(Q_{IR}) - aQ_{IR}^2 Q_P.$$

In equilibrium, $Q_P$ takes the value of one of the two minima of the double well potential $V_P(Q_P)$ (dashed black line in Fig. 1(b). For finite displacements $Q_{IR}$, this minimum is first displaced and then destabilized as $Q_{IR}$ exceeds a threshold value (colored lines in Fig. 1(b)). Importantly, the direction of this displacement is always pointed toward the opposite potential well and independent on the sign of $Q_{IR}$ [11].

The corresponding dynamics of the two modes for resonant periodic driving of $Q_{IR}$ by a mid-infrared light pulse is obtained by solving the equations of motion

$$\ddot{Q}_{IR} + \gamma_{IR}\dot{Q}_{IR} + \omega_{IR}^2 Q_{IR} = 2aQ_P Q_{IR} + f(t),$$

$$\ddot{Q}_P + \gamma_P \dot{Q}_P - 1/2 \cdot \omega_P^2 Q_P + c_P Q_P^3 = aQ_{IR}^2.$$

Here, $f(t) = A\sin(\omega_{IR} t) \cdot \exp(-4\ln 2 \cdot t^2/T^2)$ is the driving pulse with Gaussian envelope of duration $T$ and frequency $\omega_{IR}$. As displayed in Fig. 1(c), the driven mode $Q_{IR}$ experiences oscillations about its equilibrium position, whereas the ferroelectric mode $Q_P$ is subject to a unidirectional force $\partial V/\partial Q_P = aQ_{IR}^2$ [12,13].

Two regimes are found for the response of the ferroelectric mode, shown in Fig. 1(d). For small oscillation amplitudes in $Q_{IR}$, the energy minimum of the double well potential undergoes a prompt displacement, launching oscillations of $Q_P$ around a new position. In this case, the ferroelectric polarization is reduced but retains at all times the same sign of the equilibrium state. For oscillation amplitudes of $Q_{IR}$ in excess of a threshold, the model predicts that the original energy minimum is destabilized and $Q_P$ is moved to the opposite potential well, switching the polarization. A similar conversion of driven oscillatory motions into directional displacements has been previously demonstrated in a number of non-ferroelectric materials [14,15,16].

Here, we seek to validate these ideas experimentally for the case of LiNbO$_3$, a rhombohedrally distorted perovskite ferroelectric with Curie temperature $T_C$ = 1210°C. The structure of ferroelectric LiNbO$_3$ is displayed in Fig. 1(e). Below $T_C$, the Li and Nb sublattices move against the oxygen octahedra, inducing a stable electrical polarization along the rhombohedral (111)$_r$ direction that corresponds to the hexagonal c-axis. This structural transition involves atomic motions along three normal modes, one of which causes the ferroelectric polarization. Density functional theory calculations reveal strong anharmonic coupling of this 7.5-THz ferroelectric mode $Q_P$ to a high-frequency 19-THz mode $Q_{IR}$, just as described in the model above [11]. The atomic motions associated with these phonon modes are shown in Fig. 1(e).

We resonantly excited the 19-THz $Q_{IR}$ mode in a 5-mm thick x-cut single-domain LiNbO$_3$ crystal using 150-fs mid-infrared pulses. Figure 2(a) shows the electric field and Fourier transform of the 19-THz pump pulses, which were obtained by optical parametric amplification and difference frequency generation. The penetration depth of these mid-infrared pulses was 3.2μm. The ferroelectric polarization dynamics were measured by recording the transmitted second harmonic (SH) signal generated by a 35-fs, 800-nm wavelength probe pulse, which was delayed in time with respect to the excitation. As the detected SH signal was generated in the first 1.3μm of the crystal, the probed volume was homogeneously excited (see Supplementary Information Section 1 for details).

Time-resolved measurements of the SH intensity $I_{SH} \propto \left| \varepsilon_0 \chi^{(2)} \cdot E^2_{Probe}(t) \right|^2$ yielded the magnitude of the time dependent second order susceptibility $\chi^{(2)}$, which is directly

proportional to the polarization $P_s$ [17]. Higher order contributions near the second harmonic frequency $2\omega_{probe}$, most notably those descending from cubic $\chi^{(3)}$ nonlinearities at frequencies $2\omega_{probe} + \omega_{MIR}$ and $2\omega_{probe} - \omega_{MIR}$, were filtered spatially as described in Supplementary Section 2.

The time-resolved second harmonic intensity is plotted in Fig. 2(b), normalized to its equilibrium value. For excitation at fluences F < 50 mJ/cm$^2$, we observed a reduction in SH intensity to a finite value within approximately 200 fs. As predicted by the model presented above, we observed a rapid exponential recovery and coherent oscillations at 16 THz, which we attribute to a phonon-polariton mode [18,19] associated with the driven $Q_{IR}$ phonon.

For fluences above a threshold value of 60 mJ/cm$^2$ and up to the maximum fluence possible in our setup (95 mJ/cm$^2$), the second harmonic intensity was observed to vanish completely, recover to a finite value, and then vanish again, before relaxing back to the equilibrium state.

To derive, whether the ferroelectric polarization was reversed during this dynamics, we measured the time-dependent phase of the SH electric field by interfering it with a reference SH pulse, generated in a non-excited crystal as sketched in Fig. 3(a). The resulting pattern, which consisted of interference fringes on top of a Gaussian background, was detected with a charge coupled device (CCD) camera. For further analysis, this background was subtracted (see Supplementary Information Section 2). Changes in the phase of the second harmonic signal generated in the driven LiNbO$_3$ crystal appeared as changes in the spatial position of these fringes on the camera.

Figure 3(b) displays the time dependent interference signal integrated along the fringe direction for the maximum pump fluence of 95 mJ/cm$^2$. After excitation, the SH intensity first reduced with a constant phase. As the intensity reached zero at 0 fs time delay, the phase of the second harmonic field flipped by 180°, as revealed by a sudden sign change of the interference fringes. This sign change is clearly visible in Fig. 3(c), in which the interference fringes are shown for a negative delay at –200 fs and in the reversed state at +80 fs. The phase then remained constant until the SH intensity vanished again at 200 fs time delay, when the phase switched back to the initial value. Thus, for time delays between 0 fs and 200 fs, the polarization was transiently reversed.

A deeper understanding of the reversal process was obtained by analyzing the time dependent changes in SH intensity for different probe polarizations. As shown in Figure 4, the polarization dependence of the second harmonic retained the same symmetry and shape of that observed at equilibrium for all time delays. Hence, the dynamical reversal occurs only along the hexagonal *c*-axis with no rotations in the plane, consistent with atomic displacements along the ferroelectric mode $Q_P$ as described by the model above.

We further tested a second key element of the model, according to which the force acting on $Q_P$ depends on the initial equilibrium polarization state. For a polarization pointing "up", the force acts toward the "down" state, and viceversa [11]. Polarization reversal should thus be observed starting from both equilibrium states without changing the pump pulse characteristics. In Figure 5 we present the normalized amplitude (a) and phase (b) of the time dependent ferroelectric polarization measured from both initial polarization states, as extracted from the second harmonic intensity and phase according to $P_S(\tau)/P_{S,0} = \sqrt{I_{SH}/I_{SH,0}}\, e^{i\varphi(\tau)} + c.c.$. As predicted, the sign of the polarization could be transiently reversed, with a similar dynamical evolution in the two cases.

According to the model, the ferroelectric polarization reversal should further exhibit a threshold. Figure 6(a) shows a fluence dependence of the polarization at the peak of the time-resolved signal, normalized to its equilibrium value $(P_S(\tau)/P_{S,0})_{\text{peak}}$, for different excitation frequencies. Indeed, we find that for excitation pulses resonant with the high-frequency $Q_{IR}$ mode transverse optical frequency $\omega_{IR,TO}$ of 19 THz, the polarization reverses only for fluences that exceed 60 mJ/cm$^2$.

We also find a substantial increase in threshold fluence for pump pulses detuned from the resonance condition. This effect is clearly captured in Figure 6(b), where we plot the slopes of linear fits to the fluence dependence of $(P_S(\tau)/P_{S,0})_{peak}$, as shown in Figure 6(a), for different pump pulse frequencies. This susceptibility closely follows the extinction coefficient of LiNbO$_3$, and peaks at $\omega_{IR,TO}$, which is again in agreement with the above model of anharmonic coupling between the driven mode $Q_{IR}$ and the ferroelectric mode $Q_P$.

Hence, the reversal without rotation, the bi-directional nature of the effect, the existence of a switching threshold and the resonance with the frequency of the infrared mode, all indicate that the essence of this phenomenon is well understood in terms of anharmonic phonon coupling and directional displacement of the ferroelectric mode.

However, some issues require further investigation. The reversed polarization was found to be only 40% of its equilibrium value, which might be due to the limited fluence of 95 mJ/cm$^2$ available for 19 THz pump pulses with our optical setup. This effect is not predicted by the minimal model and may be a sign of spatially inhomogeneous reversal. We envisage new experiments that make use of ultrafast diffuse x-ray scattering with Free Electron Lasers to quantify the possible inhomogeneity [20].

A second issue concerns the rapid return of the polarization to the initial state. In the model, switching to a stable state is expected when the threshold is exceeded. One explanation for this discrepancy may be that only coupling to the ferroelectric mode $Q_P$ is considered in our model. As described above, the equilibrium paraelectric-ferroelectric transition involves motion along three phonon modes. The crystal structure thus might have to relax along these two additional modes in order to permanently stabilize the reversed state. As the transient reversed state has a short lifetime, this relaxation might not occur here. Thus, permanent polarization reversal on femtosecond timescales may require selective excitation of additional vibrational modes with separate pulses to facilitate the relaxation. In the Supplementary Information Section 3 we elaborate on this possibility, considering the contribution of more anharmonic terms to the lattice potential.

The rapid return to the initial polarization may also be explained by the formation of uncompensated charges after polarization reversal of only a fraction of the material. The use of thin films or other types of fabricated structures that isolate the reversed volume may improve the lifetime or clarify these points. As similar anharmonic coupling terms have been calculated for other perovskite transition-metal ferroelectrics [11] such as $PbTiO_3$ or $BaTiO_3$, the same technique could also be applied to materials with lower coercive fields.

Despite these limitations, which at the present stage make it impossible to build a non-volatile ultrafast memory unit, immediate applications for the transient polarization reversal demonstrated here can already be envisaged. For example, the ability to reverse the polarization on femtosecond timescales may be used to control complex functionalities dynamically in materials that exhibit more than one coupled ferroic order [21,22,23] or to transfer charges at ultrafast speeds across hetero-interfaces [24,25].

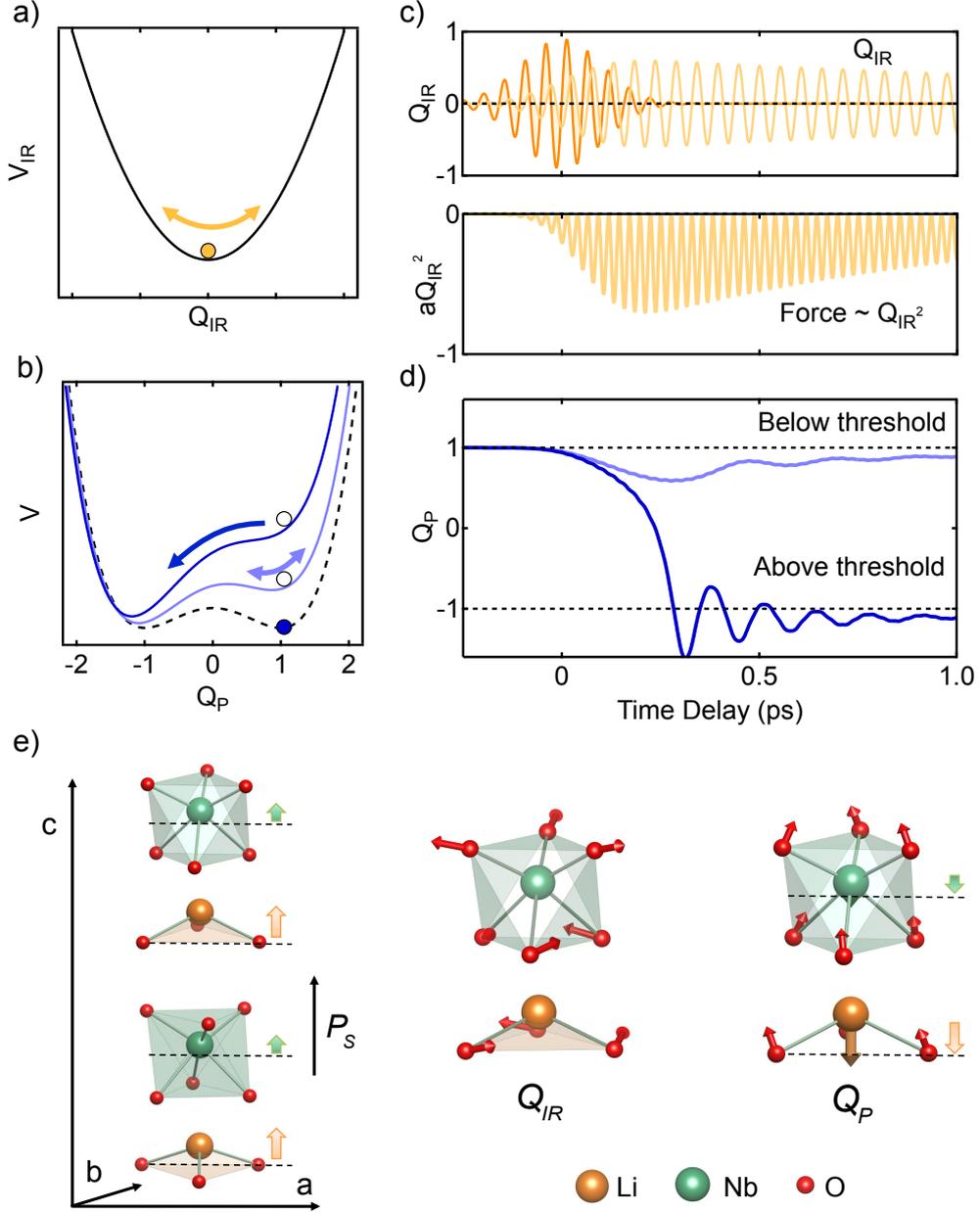

**FIG. 1**. **a)** Harmonic energy potential $V_{IR}(Q_{IR})$ of the resonantly excited high frequency mode $Q_{IR}$. **b)** Double well potential $V(Q_P, Q_{IR})$ along the ferroelectric mode $Q_P$. For finite amplitude of $Q_{IR}$, the energy minimum of the equilibrium potential ($V_P(Q_P)$, dashed line) is first displaced and then destabilized as the amplitude of $Q_{IR}$ exceeds a threshold (colored lines). **c,d)** Solution to the corresponding coupled equations of motion (see text). Following excitation of $Q_{IR}$ with mid-infrared pulses (orange), a force $F \propto Q_{IR}^2$ displaces the ferroelectric mode $Q_P$ toward lower values. For above-threshold excitation of $Q_{IR}$, the polarization reverses permanently. **e)** Crystal structure of ferroelectric LiNbO$_3$. Drawn are the hexagonal crystal axes. The Li an Nb sublattices are shifted against the oxygen octahedra (orange and green arrows), inducing a ferroelectric polarization $P_S$ along the *c*-axis. Further shown are schematics of the motions associated with the phonon modes $Q_{IR}$ and $Q_P$. The ferroelectric mode $Q_P$ involves *c*-axis motions of the Nb and Li-sublattices against the oxygen octahedra (orange and green arrows) and modulates the ferroelectric polarization.

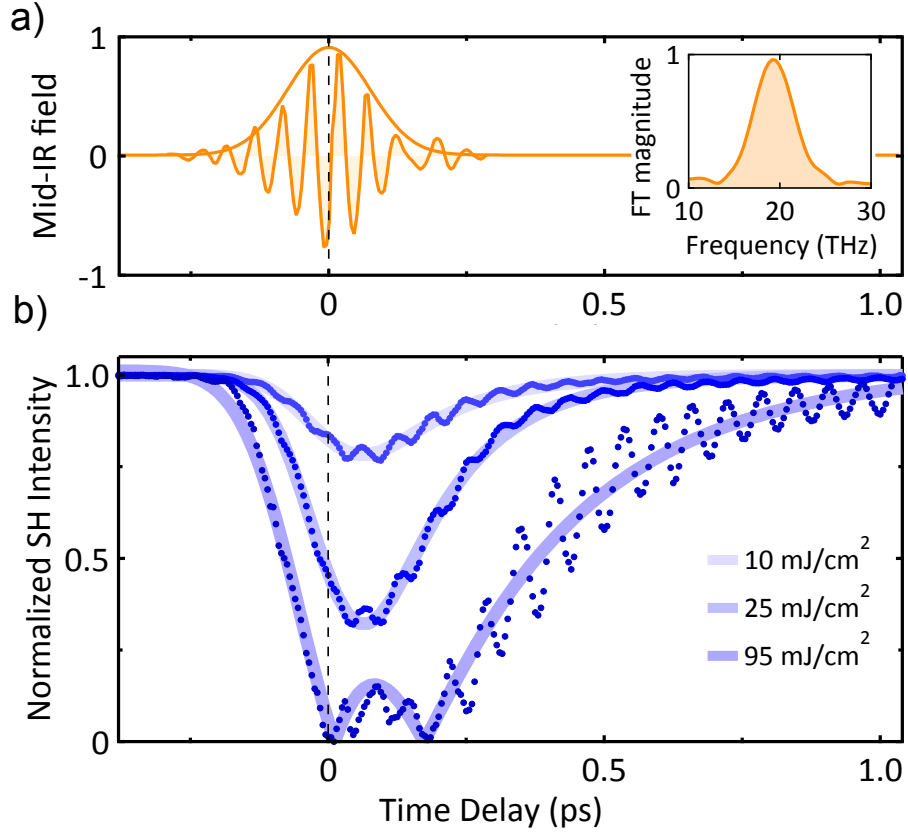

**FIG. 2**. **a)** Carrier envelope phase stable 19-THz mid-infrared pulses were used to excite the high frequency mode $Q_{IR}$ shown in Fig. 1e). The inset shows the Fourier transformation of the pulse. **b)** Time-resolved second harmonic intensity, normalized to its value before excitation. Following excitation, the second harmonic intensity reduces before relaxing back. This drop increases with pump fluence. For all fluences above 60 mJ/cm$^2$, the second harmonic signal vanishes, followed by a transient recovery before relaxing back to its equilibrium value.

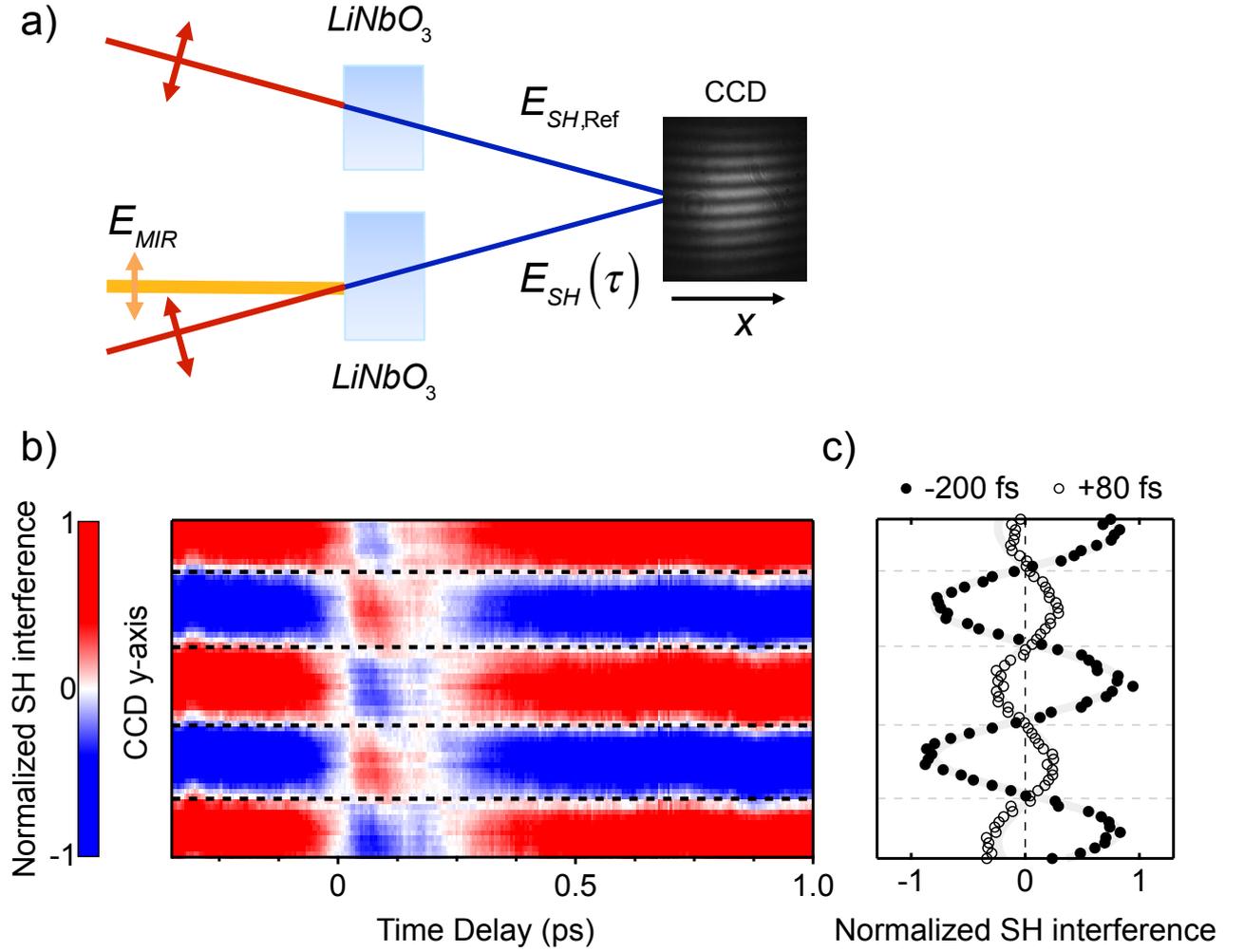

**FIG. 3**. **a**) Experimental geometry for phase sensitive measurements of the second harmonic. The time dependent second harmonic signal from the excited LiNbO$_3$ crystal was interfered with a reference second harmonic field from an unexcited sample and measured with a CCD camera. The resulting interference pattern consisted of fringes on top of a Gaussian background, which was subtracted. **b**) Time resolved measurement of the interference fringes, integrated along the x-axis of the camera and normalized. The data shows a transient phase change by 180° (sign reversal) between 0 fs and 200 fs. **c**) Normalized interference fringes at -200 fs and in the reversed polarization state at +80 fs.

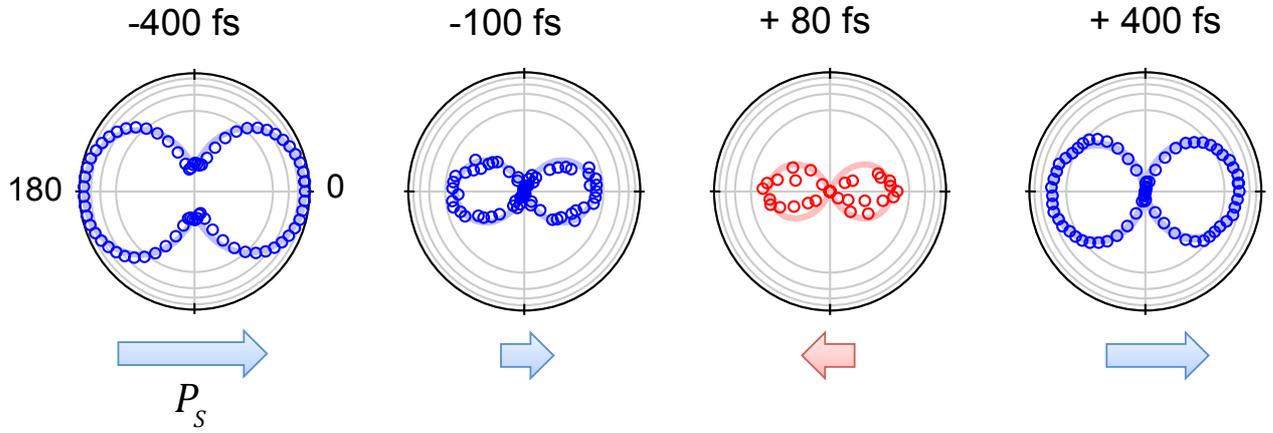

**FIG. 4**. Probe pulse polarization dependence of the second harmonic signal at four time delays after excitation. The angles 0° and 90° denote a polarization of the 800nm pulses along and perpendicular to the spontaneous polarization. The arrows indicate the time dependent amplitude and sign of the ferroelectric polarization for each time delay, determined from the measurements of Figure 3. The solid lines are fits to the data using the $\chi^{(2)}$ tensor of LiNbO$_3$.

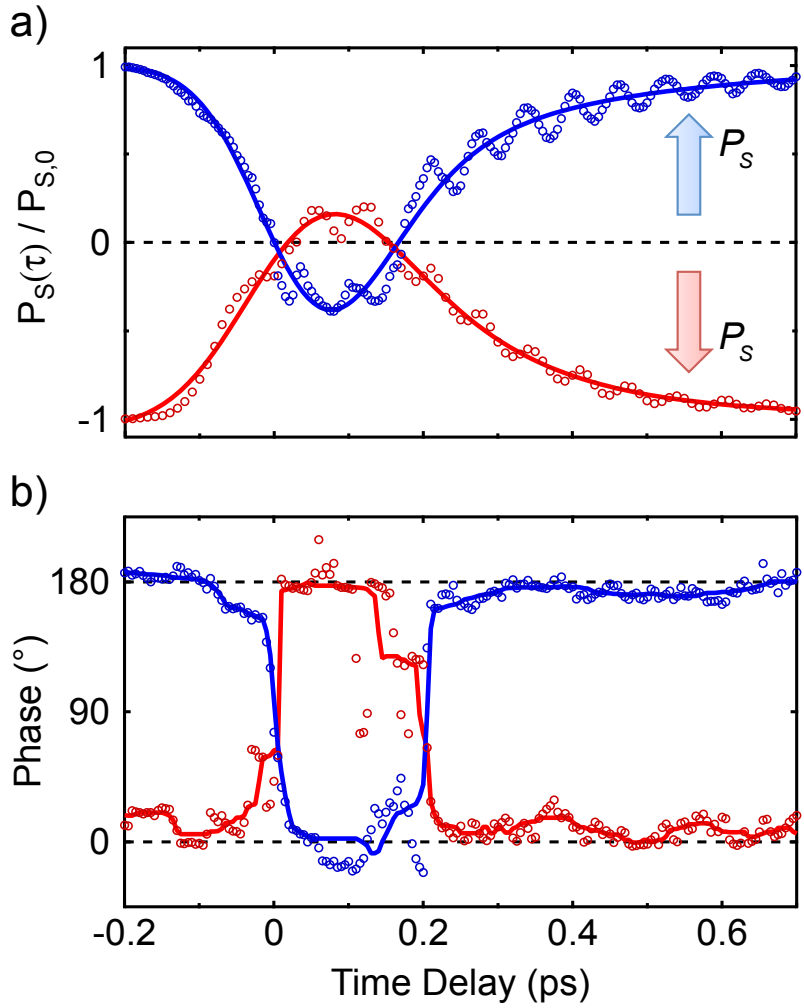

**FIG. 5**. **a)** Time-resolved normalized ferroelectric polarization $P_S(\tau)/P_{S,0}$ following excitation of LiNbO$_3$ with two opposite initial polarization states, as determined from the measurement of the SH intensity and the phase of the SH field. **b)** Phase of the SH field derived from a sine fit to the interference pattern (see text).

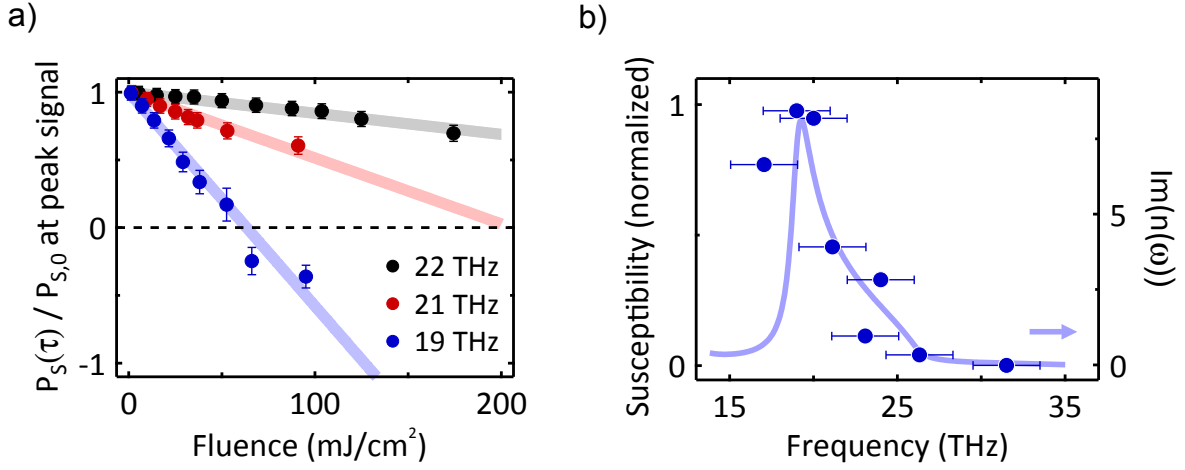

**FIG. 6**. **a)** Normalized polarization at the peak of the time dependent signal $(P_S(\tau)/P_{S,0})_{peak}$ as function of pump fluence. The three curves correspond to excitation with pulses at different frequencies. The polarization reverses for excitation above 60 mJ/cm² with 19-THz pulses, resonant with the $Q_{IR}$ mode transverse optical frequency $\omega_{IR,\,TO}$. **b)** Pump pulse frequency dependence of the susceptibility to the excitation (blue dots, left axis), defined as the slope of linear fits to the fluence dependence of $(P_S(\tau)/P_{S,0})_{peak}$, shown in Fig. 6(a). The solid line shows the extinction coefficient of LiNbO$_3$ for comparison (right axis).

**Acknowledgements:**

We thank Michael Fechner, Roberto Merlin and Alaska Subedi for valuable discussions. The research leading to these results received funding from the European Research Council under the European Union's Seventh Framework Programme (FP7/2007-2013)/ERC Grant Agreement no. 319286 (QMAC).

This work has been supported by the excellence cluster 'The Hamburg Centre for Ultrafast Imaging - Structure, Dynamics and Control of Matter at the Atomic Scale' of the Deutsche Forschungsgemeinschaft.

**Supplementary material**

**1 - Second harmonic probing from femtosecond pulses**

The sample was a 5 mm thick LBO crystal. The 1/e penetration depth of the electric field of the 19 THz pump electric field, convolved with the spectral bandwidth of 5.5 THz FWHM, was 3.2 µm. In order to trace the polarization dynamics in the excited volume, we detected non phase-matched second harmonic generation from 35-fs 800nm probe pulses, generated in a thin layer below the surface.

This second harmonic signal is the homogenous solution to the wave equation, which results from the discontinuity of the optical properties at the boundary of the crystal [26,27]. The generation occurs in a layer below the surface with thickness given by the SH coherence length of $l_c = 1.27 \mu m$, over which the dipole emission builds up constructively [26,28,29]:

$$l_c = \frac{\pi}{|k_{2\omega} - 2k_\omega|} = \frac{\lambda}{4|n_{2\omega} - 2n_\omega|}.$$

This SH pulse, which reflects the polarization dynamics in the pumped volume, propagates with a group velocity determined by the linear optical properties and dispersion of the crystal at $2\omega$.

In addition to this SH signal, a second pulse with wave vector $2k_\omega$ is generated by the fundamental as it propagates through the crystal. This inhomogeneous solution to the wave equation co-propagates both in time and space with the fundamental pulse, therefore with an apparent group velocity determined by the dispersion at $\omega$. Thus, this SH pulse separates both spatially and temporally from the interface-generated SH pulse [27,30] and its intensity is not affected by changes in the excited volume close to the surface (see Figure 1).

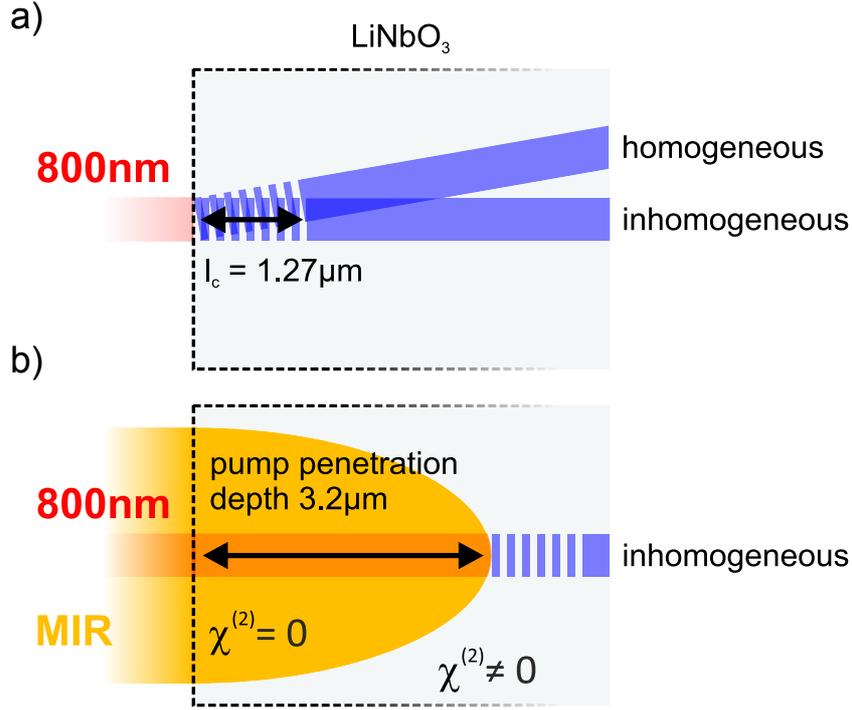

**Fig. 1 a)** Non phase-matched SH generation in bulk LiNbO$_3$. Two second harmonic signals are generated by the 800nm probe pulses, which separate spatially. The homogeneous solution to the wave equation is generated in a 1.27μm thin layer below the surface. **b)** As the quadratic nonlinearity $\chi^{(2)}$ is depleted within the excited volume, the signal of the homogeneous solution vanishes. In contrast, the SH signal of the inhomogeneous solution is generated in the bulk of the crystal, which is not excited, leaving its intensity unaffected.

These two pulses were separately detected by cross-correlating the total second harmonic radiation emitted from the 5mm thick LiNbO$_3$ crystal, with a synchronized 800-nm pulse in a $\beta$-BBO crystal. The measured intensity of the sum frequency is shown in Figure 2 together with a simulation using the SNLO software package (black curve) [31]. The homogeneous solution is delayed by 10.5 ps with respect to the inhomogeneous solution, in good agreement with the simulations. The difference in the relative amplitudes of the homogeneous and inhomogeneous solutions is due to their spatial separation on the detection crystal.

The red curve in Figure 2 shows the same measurement with the sample being excited by the 19 THz MIR pulse at a fixed time delay with respect to the 800 nm pulse. The inhomogeneous solution is almost unaffected due to the limited penetration depth of the MIR pump pulses. In contrast, the homogeneous solution vanishes due to a complete suppression of the optical nonlinearity $\chi_2$ in the excited volume close to the surface. Hence, changes in the excited surface region can be followed by measuring the SH pulses of the homogeneous solution in transmission through the bulk sample.

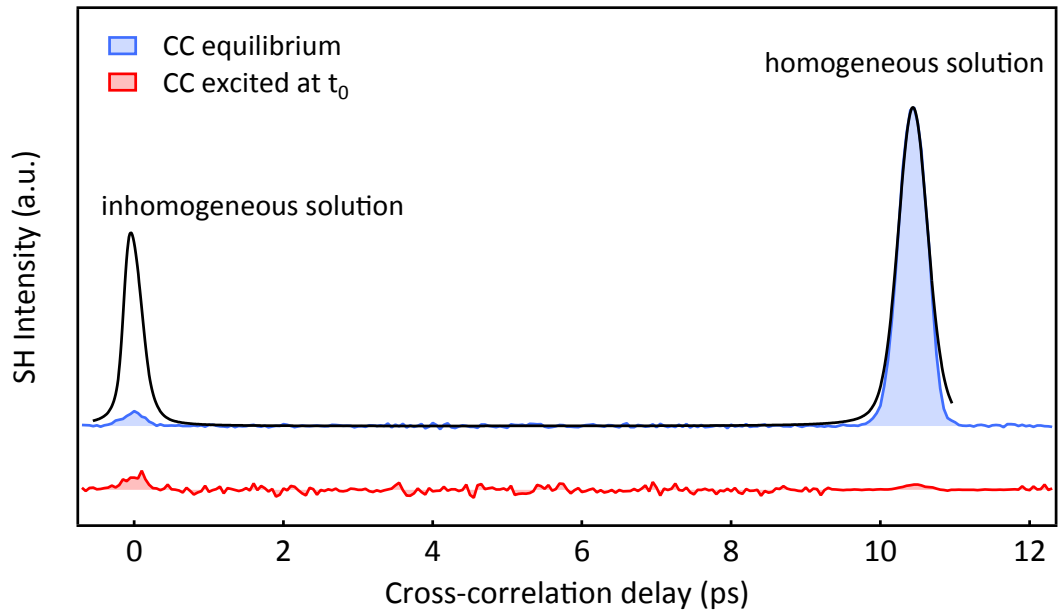

**Fig. 2** Cross-correlation between the SH light emitted from the LiNbO$_3$ crystal and a synchronized 800-nm pulse (blue curve). The homogeneous solution to the wave equation, generated at the crystal surface, is delayed by 10.5 ps with respect to the inhomogeneous solution in good agreement with simulations (black line). The excitation of the sample fully suppresses the homogeneous solution while leaving the inhomogeneous unaffected (red curve).

## 2 - Experimental Setup

**Second harmonic intensity measurement**

Figure 3 shows the non-collinear pump-probe geometry used in our experiments. An angle of 30° between MIR pump and probe beams was chosen to spatially separate the second harmonic electric fields generated due to quadratic $\chi^{(2)}$ or cubic $\chi^{(3)}$ nonlinearities. The latter process generates a SH field at frequencies $2\omega_{probe} \pm \omega_{MIR}$ by four-wave mixing with the MIR pump pulse, described by the nonlinear polarization

$$P(t) = \varepsilon_0 \chi^{(3)} E_{Probe}^2(t) \cdot \left( E_{MIR}(t) + c.c. \right).$$

As this interaction involves a momentum transfer, the third-order contribution separates spatially from the second-order contribution.

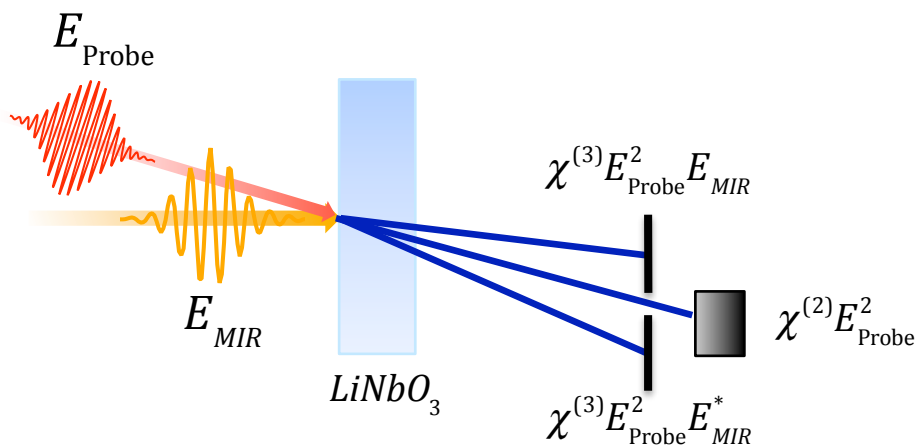

**Fig. 3** Sketch of the measurement geometry. The higher order four-wave mixing terms are separated spatially and blocked by a pinhole. The intensity of the remaining second harmonic is then detected with a photomultiplier tube.

**Second harmonic phase measurement**

The above experimental setup only allows for measuring pump induced changes in the intensity of the SH signal, which holds information on the magnitude of the ferroelectric polarization. To determine the sign of the polarization, we measured the phase of the emitted

SH electric field from the sample at all time delays by interfering it with the SH light generated in a second LiNbO$_3$ crystal. A sketch of the experimental setup is presented in Figure 3a of the main text, showing the measurement of the interference pattern on a CCD-camera. The intensity of an interference fringe at time delay $\tau$ between pump and probe pulses is given by

$$I_{SH}(\tau) = \left| E_{SH,Ref} + E_{SH,0}(\tau)e^{i\varphi(\tau)} \right|^2 = I_{SH,Ref} + I_{SH}(\tau) + 2E_{SH,Ref} \cdot E_{SH,0}(\tau)e^{i\varphi(\tau)} + c.c.,$$

where $E_{SH,Ref}$ denotes the amplitude of the SH electric field from the reference LiNbO$_3$ crystal and $E_{SH,0}(\tau)e^{i\varphi(\tau)}$ the time-delay dependent amplitude and phase of the SH field from the sample. The interference pattern can be extracted by subtracting the contribution from the terms $I_{SH,Ref} + I_{SH}(\tau)$, which appear as a Gaussian background along the x and y direction of the CCD. A sine-fit to the residual yields $\varphi(\tau)$ as shown in Figure 4.

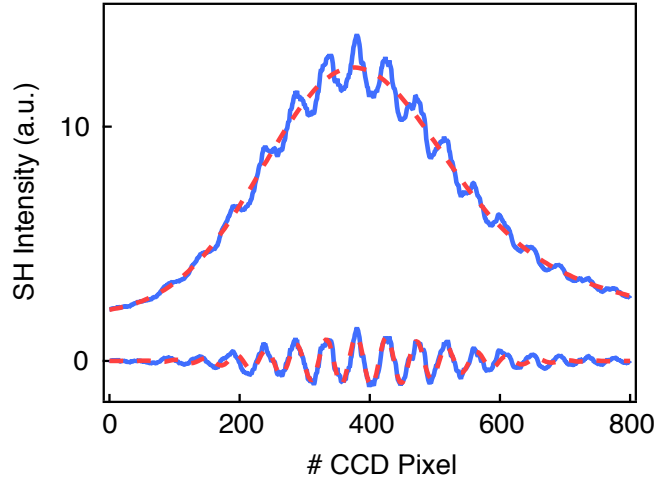

**Fig. 4** The time-delay dependent phase of the SH electric field from the sample is obtained by subtracting the Gaussian background from the measured interference pattern and fitting the residual with a sine.

## 3 - Models of the lattice dynamics

In the following section, we compare the model discussed in the manuscript and an extended model, which takes into account also cubic lattice anharmonicities along $Q_P$, as calculated in Ref. 11. Considering only the dominant coupling term $aQ_{IR}^2 Q_P$, the response of the crystal lattice to the excitation can be described by the lattice potential energy

$$V = \frac{1}{2}\omega_{IR}^2 Q_{IR}^2 + \frac{1}{2}\omega_P^2 Q_P^2 + \frac{1}{3}b_P Q_P^3 + \frac{1}{4}c_P Q_P^4 - aQ_{IR}^2 Q_P.$$

Here, $(\omega_{IR}, Q_{IR})$ and $(\omega_P, Q_P)$ denote frequencies and normal coordinates of the directly excited and the nonlinearly coupled ferroelectric mode, respectively. The constants $b_P$ and $c_P$ describe the equilibrium lattice anharmonicities.

The left panels of Figure 5 show the potential energy $V$ along $Q_P$ for both the extended model and the minimal model discussed in the main text. At equilibrium, the double well potential exhibits two stable minima, corresponding to the two polarization states ("up" and "down") (Fig. 5(b) black dashed line). The extended model only displays one minimum (Fig. 5(c), black dashed line). In both models, large amplitude oscillations of the driven mode $Q_{IR}$ dynamically induce a new energy minimum with polarization opposite to that of the initial state. Unlike the minimal model without cubic term, the extended model predicts a return of the polarization to the initial value as soon as the coherent oscillations of $Q_{IR}$ decay. This is also reflected in the dynamical response of the modes to the excitation, which is described by the equations of motion

$$\ddot{Q}_{IR} + \gamma_{IR}\dot{Q}_{IR} + \omega_{IR}^2 Q_{IR} = 2aQ_P Q_{IR} + f(t),$$

$$\ddot{Q}_P + \gamma_P \dot{Q}_P + \omega_P^2 Q_P + b_P Q_P^2 + c_P Q_P^3 = aQ_{IR}^2.$$

A solution to these equations for both the minimal and the extended model is shown in the panels on the right side of Fig. 5. The minimal model predicts permanent polarization reversal as the oscillation amplitude of the mode $Q_{IR}$ exceeds a certain threshold (Fig. 5(b)).

For the extended model polarization reversal is unstable and relaxes back to the initial polarization (Fig. 5(c)). Note that the actual physical situation is not described by either of these models as the coupling to other modes, which is needed to stabilize either polarization state along $Q_P$ are not included explicitly.

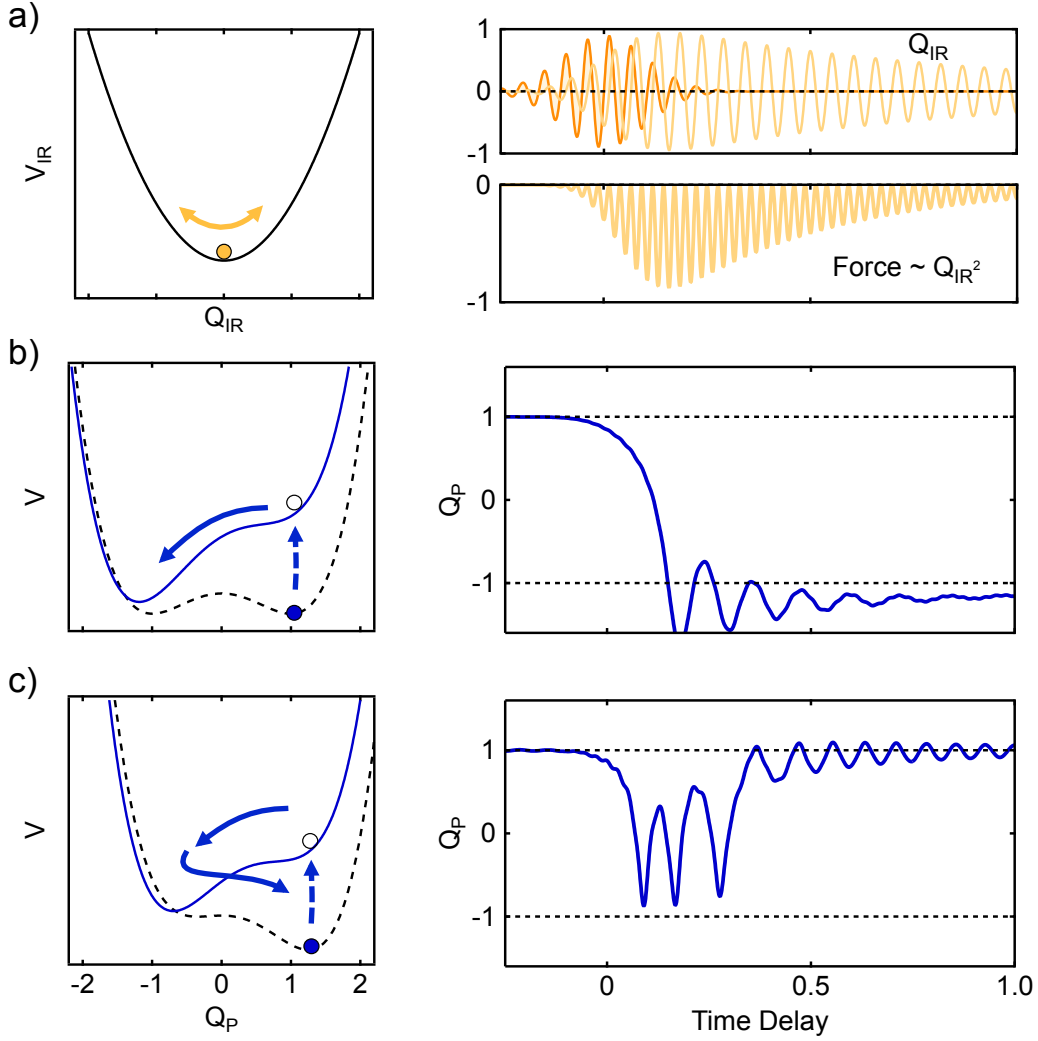

**Fig. 5 a)** Left: Parabolic energy potential of the resonantly driven infrared-active mode $Q_{IR}$. Right side: Upon excitation with MIR pulses, the atoms start to oscillate in the potential minimum. As described in the main text, due to the coupling term $aQ_{IR}^2 Q_P$ a unidirectional force $\partial V/\partial Q_{IR} = aQ_{IR}^2$ displaces the atoms along $Q_P$ toward the state of opposite polarization. **b)** For the minimal model described in the main text, excitation of $Q_{IR}$ to amplitudes above a certain threshold destabilizes the potential minimum, driving the atoms towards the stable state of opposite polarization. **c)** In case of the extended model, the reversed polarization state is only dynamically stabilized as long as IR-active mode $Q_{IR}$ oscillates at large enough amplitude.